\documentclass[conference]{IEEEtran}
\IEEEoverridecommandlockouts
\usepackage{cite}
\usepackage{amsmath,amssymb,amsfonts}
\usepackage{mathtools}
\usepackage{algorithm}
\usepackage{algpseudocode}
\usepackage{graphicx}
\usepackage{textcomp}
\usepackage{xcolor}
\usepackage{tikz}
\usepackage{pgfplots}
\usepackage{booktabs}
\usepackage{todonotes}
\usepackage{placeins}
\usepackage{tablefootnote}

\usetikzlibrary{matrix,shapes,arrows,positioning,chains}
\usetikzlibrary{calc}
\usetikzlibrary{backgrounds}
\usetikzlibrary{shadows,arrows}
\usetikzlibrary{decorations.pathreplacing}
\usetikzlibrary{decorations.markings}
\usetikzlibrary{positioning,shapes}
\usetikzlibrary{fit}
\usetikzlibrary{calc}

\pgfplotsset{compat=1.17}

\tikzstyle{vecArrow} = [thick, decoration={markings,mark=at position
   0 with {\arrow[semithick, rotate=180]{open triangle 60}},mark=at position
   1 with {\arrow[semithick]{open triangle 60}}},
   double distance=1.4pt, shorten >= 5.5pt, shorten <= 5.5pt,
   preaction = {decorate},
   postaction = {draw,line width=1.4pt, white, shorten >= 4.5pt, shorten <= 4.5pt}]
\tikzstyle{innerWhite} = [semithick, white,line width=1.4pt, shorten >= 4.5pt, shorten <= 4.5pt]

\tikzstyle{buffer}=[draw, fill=gray!20, minimum width=5em, minimum height=4ex]
\tikzstyle{arbiter}=[draw, fill=gray!20, minimum width=25ex, minimum height=4ex, rotate=90]
\tikzstyle{hnode}=[draw, fill=gray!20, minimum width=50, minimum height=16]
\tikzstyle{vnode}=[draw, fill=gray!20, minimum width=77, minimum height=20, rotate=90]
\tikzstyle{cnode}=[circle, draw, fill=gray!20, minimum size=15]
\tikzstyle{layer}=[rectangle, rounded corners, draw=black, fill=gray!20, minimum size=0em, text centered, rotate=90]
\tikzstyle{init} = [pin edge={to-,thin,black}]
\tikzset{point/.style={insert path={ node[scale=2.5*sqrt(\pgflinewidth)]{.} }}}

\def\BibTeX{{\rm B\kern-.05em{\sc i\kern-.025em b}\kern-.08em
    T\kern-.1667em\lower.7ex\hbox{E}\kern-.125emX}}
\begin{document}

\title{Code-based Cryptography in IoT: \\ A HW/SW Co-Design of HQC\\
}

\author{\IEEEauthorblockN{Maximilian Schöffel}
\IEEEauthorblockA{\textit{Microelectronic Design Research Group} \\
\textit{University of Kaiserslautern}\\
Kaiserslautern, Germany \\
schoeffel@eit.uni-kl.de}
\and
\IEEEauthorblockN{Johannes Feldmann}
\IEEEauthorblockA{\textit{Microelectronic Design Research Group} \\
\textit{University of Kaiserslautern}\\
Kaiserslautern, Germany \\
feldmann@eit.uni-kl.de}
\and
\IEEEauthorblockN{Norbert Wehn}
\IEEEauthorblockA{\textit{Microelectronic Design Research Group} \\
\textit{University of Kaiserslautern}\\
Kaiserslautern, Germany \\
wehn@eit.uni-kl.de}
}

\maketitle

\begin{abstract}
Recent advances in quantum computing pose a serious threat on the security of widely used public-key cryptosystems.
Thus, new post-quantum cryptographic algorithms have been proposed as part of the associated US NIST process to enable secure, encrypted communication in the age of quantum computing.
Many hardware accelerators for structured lattice-based algorithms have already been published to meet the strict power, area and latency requirements of low-power IoT edge devices.
However, the security of these algorithms is still uncertain.
Currently, many new attacks against the lattice structure are investigated to judge on their security.
In contrast, code-based algorithms, which rely on deeply explored security metrics and are appealing candidates in the NIST process, have not yet been investigated to the same depth in the context of IoT due to the computational complexity and memory footprint of state-of-the-art software implementations.

In this paper, we present to the best of our knowledge the first HW/SW co-design based implementation of the code-based Hamming Quasi Cyclic Key-Encapsulation Mechanism.
We profile and evaluate this algorithm in order to explore the trade-off between software optimizations, tightly coupled hardware acceleration by instruction set extension and modular, loosely coupled accelerators.
We provide detailed results on the energy consumption and performance of our design and compare it to existing implementations of lattice- and code-based algorithms.
The design was implemented in two technologies: FPGA and ASIC.
Our results show that code-based algorithms are valid alternatives in low-power IoT from an implementation perspective.
\end{abstract}


\section{Introduction}
Privacy and data integrity are a key requirement in the Internet of Things (IoT).
In many applications such as industrial IoT (IIoT), medical and healthcare, online banking, and even smart homes, highly sensitive data that should not be altered or made available to the public is transmitted over the Internet.
In the vast majority of cases, the required security is provided by a combination of symmetric cryptography and Public Key Cryptography (PKC).
However, recent advances in quantum computing severely compromise the security of the State-of-the-Art (SoA) PKC.
While they are intractable on conventional computers, the underlying mathematical problems can be solved in polynomial time using Shor's Algorithms~\cite{shor1994} once large scale quantum computers become available.
This is expected to be the case by the end of this decade~\cite{mosca2018} and thus, the US NIST is currently conducting a standardization process to find new post-quantum cryptographic (PQC) algorithms.

The Key Encapsulation Mechanisms (KEMs) in the current, third round of the US NIST PQC standardization process rely on assumptions about the computational hardness of lattice-, code-, or isogeny-based problems.
Among these, the structured lattice-based algorithms are considered as most promising candidates for future standardization and for IoT applications due to their low-complexity computations.
However, the structure of the lattices used is still the subject of cryptanalysis, and the security claims of the developers remain controversial~\cite{peik2022}.

Due to the novelty of these algorithms, crypto-agility, i.e. the ability to seamlessly replace cryptographic algorithms in case that they are vulnerable, is even more important for PQC than for SoA cryptography.
Code-based algorithms are based on different, well studied security assumptions, but have a higher computational complexity and larger memory footprints than lattice-based algorithms in state-of-the-art implementations~\cite{melchor2021hamming}.
To determine if they are a viable alternative in low-power IoT environments in case that lattice-based algorithms turn out to be vulnerable, hardware implementations are essential for a conclusive evaluation and have also been requested by the US NIST~\cite{alagic2020}.

Therefore, in this work we present to the best of our knowledge the first HW/SW co-design based implementation of the code-based Hamming Quasi Cyclic KEM (HQC)~\cite{melchor2021hamming}.
Our design deploys a custom RISC-V processor and was implemented as an application-specific integrated circuit (ASIC) and field programmable gate array (FPGA). 
In summary, the new contributions of this work are:

\begin{enumerate}
    \item We provide the first ASIC implementation of a code-based KEM from the US NIST standardization process, which is fully compatible with the NIST C reference implementation.
    \item We identify the bottlenecks of the PQC algorithm and investigate for each bottleneck the best implementation method. 
    We develop and implement software optimizations, instruction set extensions, and loosely coupled accelerators and provide detailed information about their individual benefits and overhead.
    \item We compare the energy consumption, hardware requirements, and latency of our design to SoA implementations of lattice- and code-based primitives.
\end{enumerate}

The results show that our implementation is the most efficient design.
Furthermore, we show that HQC can be implemented with a similar resource utilization as lattice-based algorithms while achieving viable performance.

This paper is structured as follows. 
In Section II, we briefly introduce the working principle of KEMs in general and HQC.
In Section III, we provide an overview of the related work and of the SoA.
In Section IV, we identify and evaluate the computational bottlenecks of HQC in software.
In Section V, we present the IoT processing system and the hardware implementation of the different accelerators.
In Section VI, we compare our results with the SoA.
In Section VII, we draw a conclusion.

\section{Background}
KEMs form a public key cryptosystem that is build out of three algorithms, Key-Generation (KeyGen), Encapsulation (Encaps) and Decapsulation (Decaps).
Unlike general purpose Public Key Encryption Schemes (PKEs), KEMs are not thought to perform any application data encryption, but are designed to establish a randomly generated shared secret between communication partners in cryptographic protocols like Transport Layer Security (TLS) similar to the state-of-the-art Diffie-Hellmann Key-Exchange.
Afterwards, this shared secret is used to derive a secret key for de- and encryption of application data with fast symmetric cryptographic algorithms like Advanced Encryption Standard (AES).
KEMs are often build out of existing PKEs using transformations like the Fujisaki-Okamoto Transform.

The first code-based PKE was introduced by McEliece in 1978 and is based on the assumption that the error-correction code used is indistinguishable from random codes~\cite{mceliece1978}.
Although the original McEliece cryptosystem, which relied on Goppa codes, remains secure to this day, the method of hiding the generator matrix of the code in the public key carries a potential vulnerability.
Attempts to reduce the key size by using more structured codes than the original McEliece approach have shown that this vulnerability can be exploited to crack the cryptosystems in 0.06 seconds~\cite{faugere2010algebraic}. 

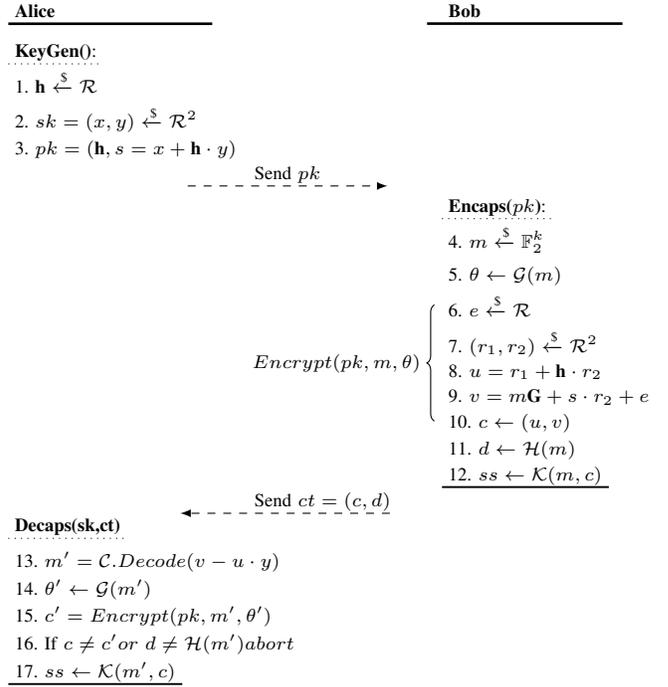
\begin{figure}[!ht]
    \centering
    \scriptsize
    \begin{tikzpicture}
    \matrix (m)[matrix of nodes, column sep=6mm, row sep=6mm, nodes={draw=none,text depth=0pt,rectangle,anchor=west}
    ]{
    \textbf{Alice} & [-13mm] & \textbf{Bob}\\ [-4mm]
    
    \textbf{KeyGen()}: & & \\ [-6mm]
    1. $\textbf{h} \xleftarrow[]{\text{\$}} \mathcal{R}$ & & \\ [-6mm]
    2. $sk=(x,y)\xleftarrow[]{\text{\$}} \mathcal{R}^2$ & & \\ [-6mm]
    3. $pk=(\textbf{h},s=x+\textbf{h}\cdot y)$ & & \\ [-6mm]
    & Send $pk$ & \\ [-5mm]
    & &\textbf{Encaps($pk$)}:\\ [-6mm]
    & & 4. $m \xleftarrow[]{\text{\$}} \mathbb{F}^k_2$ \\ [-5mm]
    & & 5. $\theta \leftarrow \mathcal{G}(m)$ \\ [-6mm]
    & & 6. $e \xleftarrow[]{\text{\$}} \mathcal{R}$  \\ [-6mm]
    & & 7. $(r_1,r_2) \xleftarrow[]{\text{\$}} \mathcal{R}^2$  \\ [-6mm]
    & & 8. $u=r_1 + \textbf{h}\cdot r_2$ \\ [-6mm]
    & & 9. $v = m\textbf{G}+s\cdot r_2 + e$ \\[-6mm]
    & & 10. $c \leftarrow (u,v)$ \\[-6mm]
    & & 11. $d \leftarrow \mathcal{H}(m)$ \\ [-6mm]
    & & 12. $ss \leftarrow \mathcal{K}(m,c)$ \\ [-6mm]
    & Send $ct=(c,d)$ & \\ [-6mm]
    \textbf{Decaps(sk,ct)} & & \\ [-5mm]
    13. $m' = \mathcal{C}.Decode(v - u\cdot y)$ & &\\ [-6mm]
    14. $\theta' \leftarrow \mathcal{G}(m')$ & & \\ [-6mm]
    15. $c' = Encrypt(pk,m',\theta')$ & & \\ [-6mm]
    16. If $c \neq c' or~d \neq \mathcal{H}(m') abort$ & & \\ [-6mm]
    17. $ss \leftarrow \mathcal{K}(m',c)$ & &\\ [-6mm]
    };
    \draw[line width=0.35mm, shorten <=-2cm,shorten >=-0cm] (m-1-1.south east)--(m-1-1.south west);
    \draw[line width=0.35mm, shorten <=-0.8cm,shorten >=2.4cm] (m-1-3.south east)--(m-1-3.south west);
    \draw[dotted] (m-2-1.south west)--(m-2-1.south east);
    \draw[dashed, shorten <=-0.8cm,shorten >=-0.8cm,-latex] (m-6-2.south west)--(m-6-2.south east);
    \draw[dotted] (m-7-3.south west)--(m-7-3.south east);
    \draw [decorate,decoration={brace,mirror,amplitude=1mm}] ($(m-10-3.west)+(-0.1,0)$) -- ($(m-14-3.west)+(-0.1,0)$) node [black,midway,xshift=-1.3cm] {$Encrypt(pk,m,\theta)$};
    \draw[line width=0.25mm] (m-16-3.south west)--(m-16-3.south east);
    \draw[dashed, shorten <=0.1cm,shorten >=-0.9cm,-latex] (m-17-2.south east)--(m-17-2.south west);
    \draw[dotted] (m-18-1.south west)--(m-18-1.south east);
    \draw[line width=0.25mm] (m-23-1.south west)--(m-23-1.south east);
    \end{tikzpicture}
    \caption{HQC KEM as defined in~\cite{melchor2021hamming} with $\mathcal{R}=\mathbb{F}_2[X]/(X^n-1)$, the hash functions $\mathcal{G}, \mathcal{H}, \mathcal{K}$, the sampling operator $\xleftarrow[]{\text{\$}}$ and the KEM's public key $pk$, private key $sk$, ciphertext $ct$ and shared secret $ss$. $\theta$ is the seed for the pseudo-random number generation during the encryption in Encaps() and Decaps().}
    \label{fig:hqc_kem}
\end{figure}

Therefore, the authors of HQC proposed a novel approach which combines two different types of codes:
\begin{enumerate}
    \item A decodable $[n,k]$ code $\mathcal{C}$ with a fixed, publicly known generator matrix $\textbf{G} \in \mathbb{F}^{k\times n}_{2}$ and the error correction capability $\delta$ based on concatenated Reed-Muller (RM) and Reed-Solomon (RS) codes.
    \item A random double-circulant $[2n, n]$ code with a publicly known parity check matrix $\textbf{h}$.
\end{enumerate}

This design rational allows HQC to use significantly smaller keys than the other code-based KEM Classic McEliece (2\,KB vs 255\,KB public key size) while still achieving the same security metrics.

Fig.~\ref{fig:hqc_kem} shows how the shared secret $ss$ is established between the communication partners using the HQC KEM.
HQC uses the Keccak-based extendable output function SHAKE as a seedexpander of a random generated seed as the scheme requires a large amount of random bytes ($n=17669$ for the smallest parameter set HQC-128).
Furthermore, the Keccak-based Secure Hash Alorithm 3 (SHA3)~\cite{nist2015} is used for the $\mathcal{G},\mathcal{H}$ and $\mathcal{K}$ functions which are required due to the KEM-DEM transformation in HQC to construct an IND-CCA2 secure KEM.

The procedure of HQC in short is as follows, a detailed description can be found in~\cite{melchor2021hamming}.
First, Alice randomly generates the parity check matrix \textbf{$h$} and the private key $sk$, from which the public key $pk$ is constructed.
Here, the polynomials $x$ and $y$ which build the secret key are hidden in the public key by multiplying \textbf{h} with $y$ and adding $x$ in $\mathcal{R}$.
Bob uses $pk$ to encrypt his randomly generated message $m$, which is the basis for the shared secret $ss$.
During this encryption, the randomly generated vectors $r1$, $r2$ and $e$ which have a fixed, predefined hamming weight are used to disguise $m$ further.
The hamming weights are selected in a way such that they still allow a correct decryption of $m$ by Alice with respect to $\delta$ with a very high probability.
The ciphertext $ct$ is sent back to Alice, who decrypts the message $m'$ and calculates $ss$ based on it.

The HQC algorithm is available in 3 different parameter sets.
This paper is focused on the NIST level 1 parameter set HQC-128.

\FloatBarrier


\section{State of the Art}
Many works have been published which deal with hardware accelerations of new PQC primitives.
Among these publications, the vast majority is focused on accelerators for lattice-based algorithms.
A cryptographic co-processor was implemented in~\cite{banerjee2019sapphire} as an ASIC to support various lattice based NIST schemes.
Fritzmann et al. developed a HW/SW based co-design on a RISCV core for the lattice-based scheme NewHope~\cite{fritzmann2019}.
In~\cite{karl2022}, FrodoKEM, an algorithm which has a high security confidentiality due to its less structured lattice, was accelerated by using a HW/SW co-design approach.

In contrast, the code-based KEMs have not yet been investigated to the same depth.
For BIKE, another code-based candidate with a comparable key size to HQC, an FPGA implementation has been proposed in~\cite{richter2022folding}.
So far, the only hardware implementation for HQC was presented by the original authors of HQC in~\cite{melchor2021hamming} and is based on FPGA HLS.
Therefore, in this work, we present the first HW/SW co-design approach of HQC and implement our design both as ASIC and FPGA.
\section{HW/SW co-design}
There are three possibilities for implementation:

\begin{enumerate}
    \item Software.
    \item Custom processor instructions.
    \item Loosely coupled accelerators.
\end{enumerate}

In a first step, the execution of the NIST reference software was profiled to determine the computational bottlenecks and the memory footprint.
In a second step, we investigated for each bottleneck the most suitable approach to find the optimum trade-off between the area, latency, memory footprint, and energy consumption.
The highest priority was assigned to software optimization, as it offers high flexibility without additional costs.
Then, if this is not efficient, custom processor instructions were considered as a second option, since they are still flexible and require little additional hardware.
Only when these two approaches were found to be ineffective a loosely-coupled accelerator was considered.

\subsection{IoT Processing System (IoT-PS)}
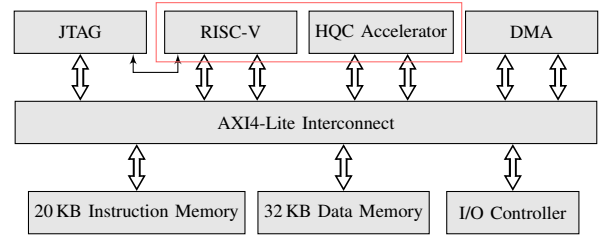
\begin{figure}[ht]
    \centering
    \scriptsize
    \begin{tikzpicture}[node distance=2.5cm,auto,>=latex']
    \node [hnode] (cpu) at (2,0) {RISC-V};
    \node [hnode] (jtag) at (0,0) {JTAG};
    \node [hnode] (mdma) at (6,0) {DMA};
    \node [hnode] (hqc) at (4,0) {HQC Accelerator};
    \node [hnode] (rom) at (0.75,-2.4) {20\,KB Instruction Memory};
    \node [hnode] (ram) at (3.5,-2.4) {32\,KB Data Memory};
    \node [hnode] (io) at (5.75,-2.4) {I/O Controller};
    \node [hnode, minimum width=220] (inter) at (3,-1.2) {AXI4-Lite Interconnect};
    \draw[vecArrow] (jtag.south) -- (inter.north-|jtag);
    \draw[vecArrow] ([xshift=10pt]hqc.south) -- ([xshift=10pt]inter.north-|hqc);
    \draw[vecArrow] ([xshift=-10pt]hqc.south) -- ([xshift=-10pt]inter.north-|hqc);
    \draw[vecArrow] ([xshift=10pt]cpu.south) -- ([xshift=10pt]inter.north-|cpu);
    \draw[vecArrow] ([xshift=-10pt]cpu.south) -- ([xshift=-10pt]inter.north-|cpu);
    \draw[vecArrow] ([xshift=10pt]mdma.south) -- ([xshift=10pt]inter.north-|mdma);
    \draw[vecArrow] ([xshift=-10pt]mdma.south) -- ([xshift=-10pt]inter.north-|mdma);
    \draw[vecArrow] (rom.north) -- (inter.south-|rom);
    \draw[vecArrow] (ram.north) -- (inter.south-|ram);
    \draw[vecArrow] (io.north) -- (inter.south-|io);
    \draw[<->] ([xshift=20pt]jtag.south) -- +(down:7pt) -- ([xshift=-20pt,yshift=-7pt]cpu.south) -- ([xshift=-20pt]cpu.south);

    \draw[red!60] ([xshift=-3pt,yshift=3pt]cpu.north west) -- ([xshift=3pt,yshift=3pt]hqc.north east) --
    ([xshift=3pt,yshift=-3pt]hqc.south east) -- ([xshift=-3pt,yshift=-3pt]cpu.south west) -- cycle;
    
    \end{tikzpicture}  
    \caption{Architecture Overview}
    \label{fig:soc}
\end{figure}

Our methodology requires a processing system that allows instruction set extensions and the efficient interfacing of loosely-coupled accelerators.
Therefore, we chose an adaptive platform that includes a RISC-V core whose instruction set architecture provides the ability to add custom instructions.
Fig.~\ref{fig:soc} shows the final architecture of the IoT-PS.
Our custom, area optimized RISC-V core supports the RV32IC instruction set which features additional compressed instructions and, therefore, significantly reduces the program size.
The Direct Memory Access (DMA) controller features a memory copy (\texttt{memcpy}) and memory initialization (\texttt{memset}) function, of which both are able to operate on byte, half-word, and word granularity.
The JTAG module provides access to the memories and the register file of the RISC-V core.
It also can be used to start, stop, and reset the IoT-PS.
Depending on the target platform, the data memory module was either based on an SRAM hard macro cell (ASIC) or Block RAM (Xilinx FPGA).
Block RAM was also used for the instruction memory in case of an FPGA implementation.
However, for the ASIC implementation we used a ROM macro cell, thus the program code is available after reset and does not need to be loaded via JTAG.
The IoT-PS features no peripheral units except the I/O controller which is used to communicate via pin toggling.

\subsection{Profiling of HQC-128}
The US NIST C reference implementation was used to identify the bottlenecks of the HQC-128 execution in our setup.
The code was compiled with optimization level 2 (O2) and simulated cycle-accurately with the RTL model of our IoT-PS.
Compared to Fig.~\ref{fig:soc}, the size of ROM and RAM had to be increased for the analysis due to the large requirements of the reference implementation.

The simulation results are shown in Table~\ref{tab:cycles_ref_impl}.
The specified clock cycles in the table refer to the processor cycles that the RISC-V core spends within the respective C function and excluding the time spent in sub-functions, e.g., gf\_mul is called during the computation of RS-Encode, but not included in its reported cycles.
For all three KEM-functions, (1) the arithmetic in $\mathcal{R}$, (2) the SHAKE-based hashing and (3) memory operations are the main contributors to the total execution time.
On top of that, the sampling operation, the RM-Decoding algorithm (4) and the finite field multiplication (5) are further contributors to the computation time.
The unsigned division, which is performed in software, is mostly used during the polynomial multiplication in (1).

\begin{table*}
    \centering
    \caption{Total cycle count and share of important functions of the NIST C Reference Implementation on the RISC-V, n.A. (not Applicable) refers to functions which are not used in this step.}
    \label{tab:cycles_ref_impl}
    \begin{tabular}{llll}
     \toprule
     \textbf{Function} & \textbf{Keygen Cycles} & \textbf{Encaps Cycles} & \textbf{Decaps Cycles} \\
     \toprule
     \toprule
     \textbf{Total} & \textbf{5609k} & \textbf{13850k} &  \textbf{19903k}\\
     \midrule
     \textbf{Arithmetic in} $\mathcal{R}$ & \textbf{1540k (27.46\%)} & \textbf{3448k (24.9\%)}  & \textbf{4989k (25.1\%)}\\
     - Vect\_Mul & 1528k & 3413k & 4942k\\
     - Vect\_Add & 12k & 35k & 47k\\
     \midrule
     \textbf{SHAKE} & \textbf{1854k (33.05\%)} & \textbf{5007k (36.15\%)} & \textbf{5414k (27.2\%)}\\
      - Keccak\_State\_Permute & 1744k & 4626k & 5005k\\
      - Keccak\_Inc\_Squeeze & 103k & 131k & 154k\\
      - Keccak\_Inc\_Absorb & 7k & 250k & 255k\\
     \midrule
     \textbf{RS-RM Code} & \textbf{n.A.} & \textbf{26k (0.18\%)} & \textbf{1440k (7.24\%)}\\
     - RS-Encode & n.A. & 26k & 26k\\
     - RS-Decode & n.A. & n.A. & 56k\\
     - RM-Decode & n.A. & n.A. & 1358k\\
     \midrule
     \textbf{Sampling} & \textbf{81k (1.44\%)} & \textbf{155k (1.1\%)} & \textbf{236k (1.18\%)}\\
     \midrule
     \textbf{Memory-Operation} & \textbf{2071k (36.92\%)} & \textbf{5068k (36.59\%)} & \textbf{7175k (36.05\%)}\\
     - memcpy & 2045k & 5021k & 7092k\\
     - memset & 26k & 47k & 83k\\
     \midrule
     \textbf{Rest} & \textbf{63k (1.12\%)} & \textbf{146k (2.17\%)} & \textbf{649k (3.26\%)}\\
     - unsigned division & 49k & 100k & 151k\\
     - gf\_mul & n.A. & 20k & 162k\\
     \bottomrule
    \end{tabular}
\end{table*}

In (1), the largest part is accounted by the multiplication of the large polynomials ($n=17669$ for HQC-128), which are represented as bit vectors. This includes the subsequent reduction by $X^n - 1$ of the intermediate result, and is performed, for example, in steps 3., 8. and 9. in Fig.~\ref{fig:hqc_kem}.
The multiplication complexity is reduced by the fact that one of the vectors is sparse and has a small, known hamming weight $w \leq 75$, which allows to consider only the non-zero coefficients in the sparse polynomial during processing.
The execution of the multiplication consists mostly of XOR operations for adding the binary coefficients of the same degrees and SHIFT / AND operations to determine the degree of the intermediate results.
Due to the high degree of the polynomials, a large number of LOAD and STORE instructions is required during the computation.

Keccak's permutation function in (2) is the computational core of the sponge construction in SHA3 and consists of bitwise AND, XOR, and rotate operations on the 25 lanes of 64 bits each. 
The major bottleneck in this permutation is the interdependence of the intermediate results which causes the contents of the processor registers to be swapped with the main memory multiple times during the execution of one of the 24 rounds.

The large overhead of the memory operations in (3) is driven by two reasons.
First, the RISC-V core supports only one outstanding memory read or write access at a time.
It waits for a slave response before continuing the program execution.
Second, the reference implementation is not optimized for low memory usage, e.g., it often stores multiple copies of temporary results, initializes a larger number of arrays, or copies parts of arrays to different memory locations.

\begin{table} [!h]
    \centering
    \caption{Stack Memory and Code-Size of the NIST C Reference Implementation of HQC-128 on our RISC-V core.}
    \label{tab:mem_ref_impl}
    \begin{tabular}{llll}
    \toprule
    & \textbf{Keygen} & \textbf{Encaps} & \textbf{Decaps} \\
    \toprule
    \textbf{Code Size} & 10.798~KB & 17.015~KB & 22.378~KB \\
    \midrule
    \textbf{Stack Memory} & 53.018~KB & 68.714~KB & 77.762~KB \\
    \bottomrule
    \end{tabular}
\end{table}

\subsection{Software Optimization}
In multiple functions, the reference implementation uses non-optimal data-types which increases the number of required memory accesses and processor instructions.
An example of this is the comparisons in Step~16 of Fig.~\ref{fig:hqc_kem}, which are performed on a byte boundary rather than a processor word boundary.
The memory footprint and computation time was further improved by removing redundant arrays which often get initialized with zeroes or are the target of \texttt{memcpy} operations.
The operations are performed with pointers instead.
For the remaining memory operations, the time required for \texttt{memcpy} and for array initialization via \texttt{memset} were accelerated by using the DMA controller of the platform.

\subsection{Instruction Set Extension}
The bottleneck of RS-Encoding and -Decoding is caused by the multiplication in $\mathbb{F}_{2^8}$.
This operation has only three operands, including the generator polynomial, and one return value with the size of one byte each.
A $\mathbb{F}_{2^8}$-Unit is added to the RISC-V core which is able to perform the operation shown in Equation~\ref{eq:gf_mul}, where $a = (a_{15},\cdots, a_0)$ and $b = (b_{7},\cdots, b_0)$ are the input operands, and $d = (d_{14},\cdots, d_0)$ is the output.
This unit is made accessible via both an R-type and an I-type custom instruction, where register \texttt{rs1} is used as operand $a$, register \texttt{rs2} respectively the immediate value \texttt{imm} is used as operand $b$, and the output $d$ is stored in register \texttt{rd}.

\begin{equation}
\begin{gathered}
    (a_{15}\cdot x^7 + \cdots + a_{8}\cdot x^0) \cdot (b_{7}\cdot x^7 + \cdots + b_{0}\cdot x^0) \\
    + (a_{7}\cdot x^7 + \cdots + a_{0}\cdot x^0) \Rightarrow (d_{14}\cdot x^{14} + \cdots + d_{0}\cdot x^0)
\end{gathered}
\label{eq:gf_mul}
\end{equation}

Using these custom instructions, a multiplication in $\mathbb{F}_{2^8}$ is performed within four clock cycles.

\begin{figure*}[ht]
	\centering
		\includegraphics[width=0.7\textwidth]{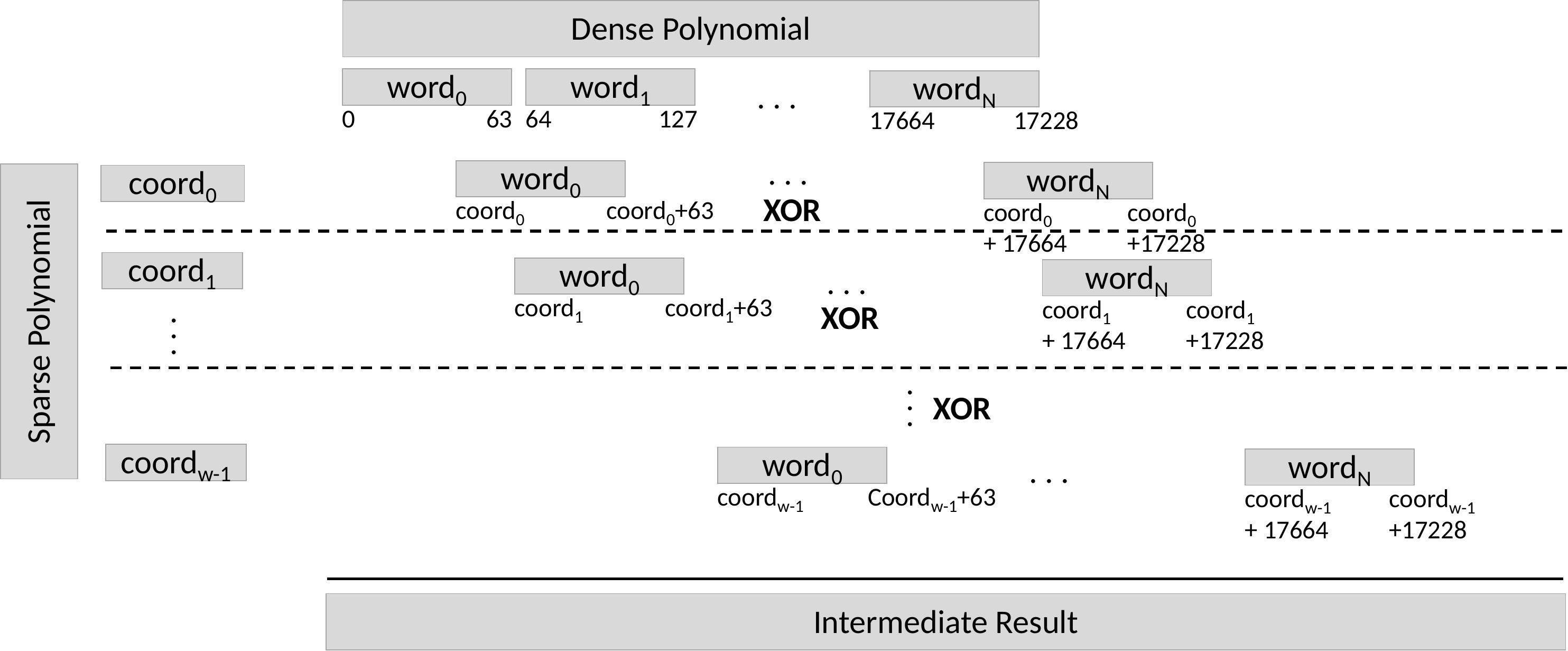}
		\caption{Working principle of the polynomial multiplication in $\mathcal{R}$ with $n=17669$ and 64-bit memory words.}
	\label{fig:gf2x_mul}
\end{figure*}

\subsection{Loosely-Coupled Accelerators}
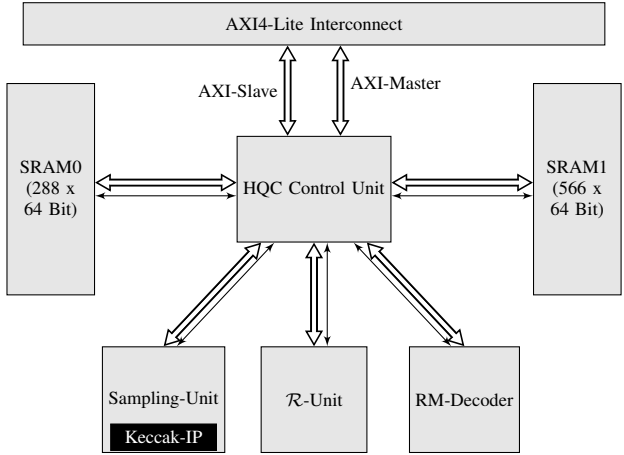
\begin{figure}[ht]
    \centering
    \scriptsize
    \begin{tikzpicture}[node distance=2.5cm,auto,>=latex']
        \node [hnode, minimum width=220] (inter) at (0,0) {AXI4-Lite Interconnect};
        \node [hnode, minimum width=40, minimum height=40] (ctrl) at (0,-2.2) {HQC Control Unit};
        \node [hnode, minimum width=10, minimum height=80,text width=1cm,align=center] (sram0) at (-3.5,-2.2) {SRAM0 (288 x 64 Bit)};
        \node [hnode, minimum width=10, minimum height=80,text width=1cm,align=center] (sram1) at (3.5,-2.2) {SRAM1 (566 x 64 Bit)};
        \node [hnode, minimum width=40, minimum height=40] (r_unit) at (0,-5) {$\mathcal{R}$-Unit};
        \node [hnode, minimum width=40, minimum height=40] (sampling_unit) at (-2,-5) {Sampling-Unit};
        \node [hnode, minimum width=40, minimum height=40] (rm_decoder) at (2,-5) {RM-Decoder};
        \node [hnode, minimum width=40, minimum height=10, fill=black, text=white] (keccak) at (-2,-5.5) {Keccak-IP};
        \draw[vecArrow] ([xshift=-10pt]ctrl.north) -- ([xshift=-10pt]inter.south) node [black,midway,xshift=-1pt] {AXI-Slave};
        \draw[vecArrow] ([xshift=10pt]ctrl.north) -- ([xshift=10pt]inter.south) node [black,right,xshift=1pt,yshift=-14pt] {AXI-Master};
        \draw[vecArrow] ([yshift=2.5pt]sram0.east) -- ([yshift=2.5pt]ctrl.west);
        \path[<->] ([yshift=-2.5pt]sram0.east) edge ([yshift=-2.5pt]ctrl.west);
        \draw[vecArrow] ([yshift=2.5pt]sram1.west) -- ([yshift=2.5pt]ctrl.east);
        \path[<->] ([yshift=-2.5pt]sram1.west) edge ([yshift=-2.5pt]ctrl.east);
        \draw[vecArrow] (ctrl.south) -- (r_unit.north);
        \path[<->] ([xshift=5pt]ctrl.south) edge ([xshift=5pt]r_unit.north);
        \draw[vecArrow] ([xshift=-20pt]ctrl.south) -- (sampling_unit.north);
        \path[<->] ([xshift=-15pt]ctrl.south) edge ([xshift=5pt]sampling_unit.north);
        \draw[vecArrow] ([xshift=20pt]ctrl.south) -- (rm_decoder.north);
        \path[<->] ([xshift=15pt]ctrl.south) edge ([xshift=-5pt]rm_decoder.north);
    \end{tikzpicture}  
    \caption{HQC hardware accelerator.}
    \label{fig:hqc_acc}
 \end{figure}

Fig.~\ref{fig:hqc_acc} shows the block diagram of the loosely-coupled HQC accelerators.
To enable parallel calculations between the processor and the accelerator, the accelerator has both an AXI slave and an AXI master interface and fetches its calculation inputs (e.g. the polynomials) from the processor's main memory via the master interface, according to the processor command which was previously received via the slave interface.
Due to the data dependencies between the steps in the HQC-KEM, and based on the previous observation that the bottleneck in the HQC is driven by memory accesses, we decided that enabling parallel read/read or read/write accesses is more beneficial than running the dedicated compute units in parallel.
Therefore, only two SRAMs are used and shared between the compute units, and only one of the compute units is processing at the same time.
The access to the SRAMs and the operation mode of the compute units are managed by one control unit.

The $\mathcal{R}$-Unit implements the addition, multiplication and reduction of the polynomials in $\mathcal{R}$.
Fig.~\ref{fig:gf2x_mul} shows the working principle of the polynomial multiplication.
The values of the sparse polynomial contain the locations of its non-zero coordinates.
In our implementation, we iterate through the multiplication by a word-by-word shift of the dense polynomial by the coordinates given in the sparse polynomial.

\FloatBarrier

After the shift, the interim result is XOR-ed with the word that is currently stored at the respective location in memory and the carry out with respect to the word alignment of the memory is calculated.
The $\mathcal{R}$-Unit is designed such that only two cycles are necessary for calculating a resulting word.
In the first cycle the address of the dense polynomial and the intermediate polynomial are calculated and read based on the coordinate, while in the second cycle the new value and the carry-out are calculated and written to memory.

The Sampling-Unit combines the sponge functions squeeze and absorb of the incremental version of SHAKE, the permutation function of Keccak and the rejection-based sampling, during which SHAKE functions are used as extendable output functions (XOF).
For the permutation function, a highspeed open-source implementation by the original authors of Keccak was used, which executes the permutation in 24 clock cycles~\cite{bert2022}.

As shown in Table~\ref{tab:cycles_ref_impl}, the vast majority of decoding time is spent on the RM-Codes, which employs a Maximum Likelihood (ML) algorithm based on the Hadamard Transform and a subsequent peak-search for the highest value in the transformed codeword.
Because HQC uses duplicated RM codes, the decoding algorithm must be preceded by another transformation function~\cite{melchor2021hamming}.
This transform requires many single-bit operations, and thus, the implementation in software is not efficient.
Therefore, we adapted the transform and the subsequent decoding steps to an efficient hardware-implementation.

\begin{table*}
    \centering
    \caption{Cycle Count of the HQC-KEM in our setup with the different hardware modules. The improvement refers to speedup with respect to the NIST C reference implementation without any hardware accelerators.}
    \label{tab:opt_impl_cycles}
    \begin{tabular}{lcccccc}
    & \multicolumn{2}{c}{\textbf{Keygen}} & \multicolumn{2}{c}{\textbf{Encaps}} & \multicolumn{2}{c}{\textbf{Decaps}} \\
    & Cycles & Improvement & Cycles & Improvement & Cycles & Improvement \\
    \midrule
    Reference                   & 5609k & -         & 13850k&           & 19903k    & -         \\
    \midrule
    DMA + SW\_OPT               & 3587k & 36.0\%    & 7044k & 49.1\%    & 10851k    & 45.5\%    \\
    + $\mathcal{R}$-Unit        & 1862k & 66.8\%    & 5183k & 62.6\%    & 7245k     & 63.6\%    \\
    + Sampling-Unit             & 1623k & 71.1\%    & 1955k & 85.9\%    & 5176k     & 74.0\%    \\
    + RM-Decoder                & n.A.  & n.A.      & n.A.  & n.A.      & 9636k     & 51.6\%    \\
    + $\mathbb{F}_{2^8}$-Instruction   & n.A.  & n.A.      & 7028k & 49.3\%    & 10722k    & 46.1\%    \\
    \midrule
    + All Modules & 56k & 98.9\% & 131k & 99\% & 557k & 97.2\% \\
    \midrule
    Code Size       & 1.5\,KB & 86.1\%    & 6.6\,KB & 61.2\%    & 13.362\,KB  & 40.3\% \\
    Stack Memory    & 10\,KB  & 81.1\%    & 24\,KB  & 65.7\%    & 31\,KB      & 60.1\% \\
    \bottomrule
    \end{tabular}
\end{table*}

\section{Results and Comparison}
\begin{figure}[!b]
	\centering
	\vspace{-4ex}
		\includegraphics[width=0.9\linewidth]{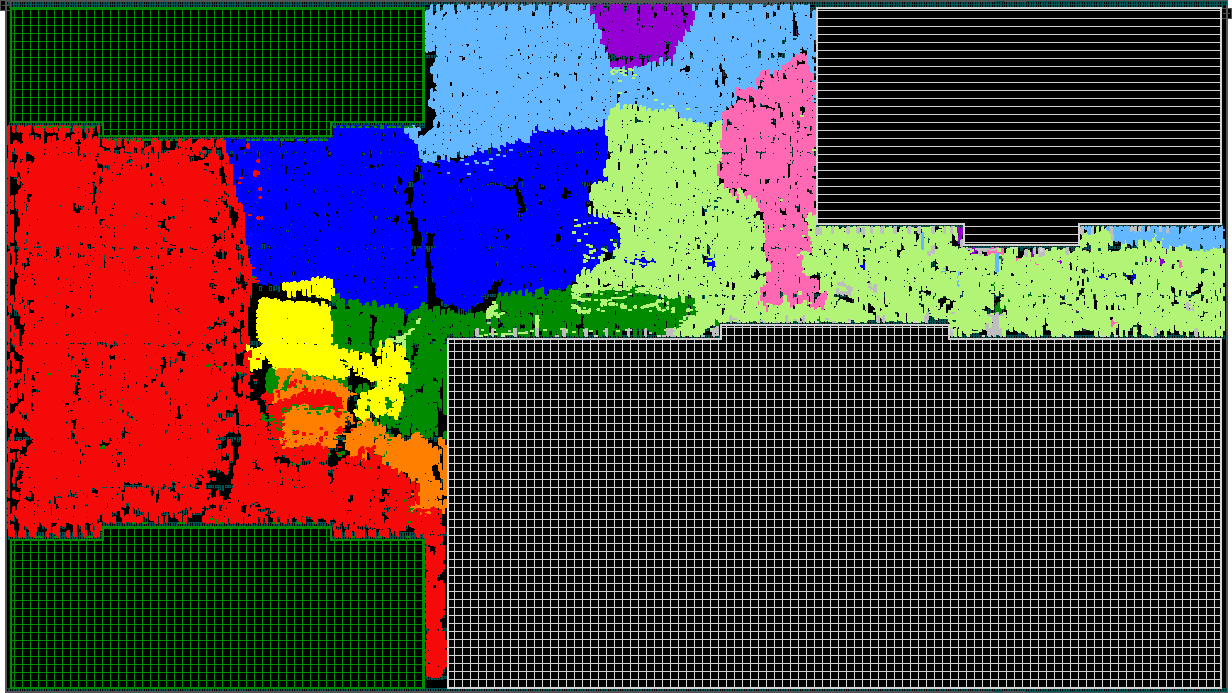}
		\caption{Layout; RISC-V - blue; JTAG - cyan; Interconnect - lime; I/O Controller - purple; DMA - pink; Sampling Unit - red; RM-Decoder - orange; $\mathcal{R}$-Unit - yellow; HQC Control Unit, SRAM0, SRAM1 - green; ROM - white, horizontal strips; RAM - white, cross pattern}
	\label{fig:asic}
\end{figure}

The IoT-PS presented was implemented in a 22\,nm FD-SOI technology from GlobalFoundries under worst case Process, Voltage and Temperature (PVT) conditions (125\,°C, 0.72\,V for timing; 25\,°C, 0.8\,V for power).
Synthesis is performed with the Synopsys DesignCompiler, Place\&Route is carried out with the Synopsys IC-Compiler. The SRAMs were generated by the INVECAS Memory Compiler.
Power numbers are calculated with back-annotated wiring data.
The layout of ASIC IP core presented in this work can be seen in Figure~\ref{fig:asic} has a size of $0.12\,mm^2$ with an aspect ratio of $1.77$ and a maximum frequency of $700\,MHz$. 
The IoT-PS presented was also implemented on a Xilinx Artix xc7a100tcsg324-3 using Xilinx Vivado for a better comparison to existing work.

\subsection{Impact of individual optimizations on the overall run time}
Table~\ref{tab:opt_impl_cycles} shows the extent to which the presented hardware modules accelerate the computation time of the three KEM functions.
As illustrated, the use of a DMA and the software optimization already provide a significant speed up between $36\%$ and $49\%$ over the reference implementation.
This shows that the NIST C reference implementation is not meant to be used in IoT devices without optimizations.
The loosely coupled Sampling-Unit is the accelerator that provides the a runtime reduction of at least $50\%$ compared to the optimized software implementation with DMA in all KEM functions.
The $\mathcal{R}$-Unit, however, reduces the runtime only between $26.5\%$ and $48\%$ depending on the KEM function.
The RM-Decoder has the least impact on the performance since it is only used in Decaps. Also it is able to achieve a runtime reduction of $11.2\%$.
The $\mathbb{F}_{2^8}$-Instructions give only a minor speedup on its own, but shows its potential in combination with all loosely coupled accelerators.

\subsection{Resource distribution of the individual hardware modules}

Figure~\ref{fig:asic} shows a qualitative area distribution of the distinct modules for ASIC, while Table~\ref{tab:fpga_util} shows a quantitative distribution for FPGA.
The $\mathcal{R}$-Unit shows the highest area efficiency among all loosely coupled accelerators.
Although this unit has a lower runtime reduction compared to the Sampling-Unit, it needs less than $10\%$ of its FPGA resources.
The RM-Decoder, however, requires fewer resources than the $\mathcal{R}$-Unit, but also offers the least performance gain and is only used in Decaps.
Therefore, it is far less efficient compared to both $\mathcal{R}$-Unit and Sampling-Unit.
The $\mathbb{F}_{2^8}$-Unit, which is used by the custom instructions, requires only negligible resources.
However, the decoding of these instructions as well as the controlling of the unit requires additional resources which are hidden inside the RISC-V core.

\begin{table}[!h]
    \centering
    \caption{Resource utilization on FPGA (Artix7)}
    \label{tab:fpga_util}
    \begin{tabular}{lrrr}
         \toprule
                                &\textbf{LUTs}  & \textbf{Registers} & \textbf{Block RAM}     \\
         \midrule           
         RISC-V             & 2210  & 1682      & 0             \\
         $^{\llcorner}$ $\mathbb{F}_{2^8}$-Unit   & 27    & 0         & 0             \\
         Interconnect           & 2775  & 1919      & 0             \\
         Memories               & 53    & 6         & 24            \\
         HQC Accelerator        & 7920  & 2370      & 3             \\
         $\vdash\mathcal{R}$-Unit      & 565   & 117       & 0             \\
         $\vdash$RM-Decoder    & 435   & 63        & 0             \\             
         $^{\llcorner}$Sampling Unit         & 5610  & 1814      & 0             \\             
         $\quad^{\llcorner}$Keccak Permute & 4685  & 1622      & 0             \\             
         DMA                    & 489   & 412       & 0             \\
         JTAG                   & 452   & 546       & 0             \\
         I/O Controller         & 41    & 68        & 0             \\
         \midrule           
         IoT-PS                    & 13934 & 7003      & 27            \\
         \bottomrule
    \end{tabular}
\end{table}

\begin{table*}[h]
    \centering
    \caption{Comparison of clock cycles, frequncy and FPGA resources for different state-of-the-art implementations. HW/SW refers to implementations based on HW/SW co-design, full refers to full hardware implementations of the respective scheme. NewHope and Kyber are structured-lattice based algorithms, FrodoKEM is based on less-structured lattices, and the remaining implementations are based on codes.}
    \label{tab:compare_performance}
    \begin{tabular}{lcccccccl}
        \textbf{Implementation} & \textbf{Keygen} & \textbf{Encaps} & \textbf{Decaps} & \textbf{Frequency} & \multicolumn{3}{c}{\textbf{FPGA Resources}} & \textbf{Target Plattform}\\
        & Cycles & Cycles & Cycles & MHz & LUTs & FFs & BRAMs & \\
        \toprule
        HQC (low area, HW)~\cite{melchor2021hamming} & 630k & 1500k & 2100k & 132 & 8.9k & 4k & 14 &FPGA (Xilinx Artix-7) \\
        HQC (low latency, HW)~\cite{melchor2021hamming} & 40k & 89k & 190k & 148 & 20k & 16k & 12.5 &FPGA (Xilinx Artix-7)\\
        BIKE (low area, HW)~\cite{richter2022folding} & 2671k & 153k & 1628k & 121 & 13k & 5k & 17 &FPGA (Xilinx Artix-7) \\
        BIKE (low latency, HW)~\cite{richter2022folding} & 259k & 12k & 189k & 96 & 53k & 7k & 49& FPGA (Xilinx Artix-7) \\
        NewHope (HW/SW)~\cite{fritzmann2019} & 357k & 590k & 167k & n.A. & 11k & 5k & 1 & FPGA (Xilinx Zynq-7000)\\
        FrodoKEM (HW/SW)~\cite{karl2022} & 23.4M & 25.5M & 25.3M & 100 & 5.6k & 1.1k & 0 & FPGA (Xilinx Zynq Ultrascale+)\\
        Kyber (HW/SW)~\cite{banerjee2019sapphire} & 75k & 132k & 142k & 72 & n.A. & n.A. & n.A. & ASIC (40nm) \\
        \midrule
        HQC on Cortex M4 (SW) & 1048k & 2436k & 4001k & 64 & n.A. & n.A. & n.A. & nRF52840 \\
        \textbf{This Work HQC (DMA+SW\_OPT)} & 3587k & 7044k & 10851k & 700 & n.A. & n.A. & n.A. & ASIC (22nm) \\
        \textbf{This Work HQC (HW/SW)} & 56k & 131k & 557k & 700 & n.A. & n.A. & n.A. & ASIC (22nm) \\ 
        \textbf{This Work HQC (HW/SW)} & 56k & 131k & 557k & 100 & 8k & 2.4k & 3 & FPGA (Xilinx Artix-7) \\
    \end{tabular}
\end{table*}

\subsection{Comparison to State of the Art}
Table~\ref{tab:compare_performance} presents the required clock cycles for the different KEM functions of SoA implementations and of our work.
We use the number of clock cycles as metric rather than absolute computation time.
This is, for our our work, a pessimistic comparison due to the high achievable clock frequency.
However, even under this assumption, our implementation requires a comparable number of clock cycles as the low-latency HQC implementation, while it requires less hardware than its low area hardware implementation.
Compared to BIKE, the other code-based KEM, our implementation requires about the same amount of clock cycles like the low latency implementation while using significantly less hardware resources.
The implementation of FrodoKEM, which would be an alternative if structured lattice-based KEMs like Kyber and NewHope are proven to be vulnerable to attacks, is overall 100 times slower than our work.

Table~\ref{tab:energy} shows the energy consumption of our implementation, of HQC on a Cortex M4 processor and of the structured lattice-based KYBER, also implemented as an ASIC.
As can be seen, our implementation requires considerably less energy than the pure software on the Cortex M4 and also less than KYBER's ASIC implementation, which, however, was implemented on a larger technology node.

\begin{table}[!hb]
    \centering
    \caption{Comparison of energy consumption. For a trade-off between power and latency, our design was implemented and simulated with a 200 MHz clock.}
    \label{tab:energy}
    \begin{tabular}{lccc}
    \toprule
    \textbf{Implementation} & \textbf{Keygen} & \textbf{Encaps} & \textbf{Decaps}\\
    & $\mu J$ & $\mu J$ & $\mu J$ \\
    \midrule
    Kyber (HW/SW)~\cite{banerjee2019sapphire} & 5.97 & 9.37 & 11.25\\
    HQC on Cortex M4 (SW) & 500 & 1184 & 1872\\
    \textbf{This Work HQC (HW/SW)} & 1.02 & 2.41 & 7.1 \\
    \bottomrule
    \end{tabular}
\end{table}

\section{Conclusion}
In this work, we investigated the performance of the code-based post-quantum KEM HQC in the context of low power IoT system.
We presented the first ASIC implementation of a code-based US NIST PQC candidate.
With a combination of software optimizations, instruction set extensions, and loosely coupled hardware accelerators, we achieve similar performance to the full SoA hardware implementation of HQC, but require significantly less hardware resources and provide more flexibility.
Compared to SoA implementations of lattice-based algorithms, we have shown that code-based algorithms are promising alternatives in IoT based applications in terms of energy efficiency, computation time, and required hardware.

\section*{Acknowledgement}
This paper was partly founded by the German Federal Ministry of Education and Research as part of the project ``SIKRIN-KRYPTOV'' (16KIS1069).
\bibliographystyle{IEEEtran}
\bibliography{references}

\end{document}